\documentstyle[12pt,psfig]{article}
\def\bge{\begin{equation}}
\def\ene{\end{equation}}
\def\bg{\begin{eqnarray}}
\def\en{\end{eqnarray}}
\def\nn{\nonumber}
\def\aleq{\stackrel{<}{\sim}}

\def\qsl{\not\!q}
\def\pslash{\not\!p}
\def\bi{\bibitem}

\begin{document}
\title{Spin Dependent Structure Functions of Nucleons and Nuclei}
\author{A.W. Thomas\\
Department of Physics and Mathematical Physics\\
University of Adelaide \\
Adelaide, S.A. 5005, AUSTRALIA}
\date{}
\maketitle
\vspace{-3.5truein}
\begin{flushright}
{\small
ADP-94-21/T161  \\
Invited paper at SPIN '94  \\
Bloomington, September 1994}
\end{flushright}
\vspace{2.5truein}
\begin{abstract}
We review recent progress
 in the understanding of the spin structure of the nucleon.  For the 
free nucleon the issues addressed include the status of the Bjorken and 
Ellis-Jaffe sum-rules and the role of the axial anomaly.  We outline 
recent work connecting the quark models familiar from hadron 
spectroscopy to the spin and flavour dependence of the parton distributions.  
Finally we review the current understanding of nuclear spin structure 
functions and particularly the extraction of the neutron spin 
structure function from deuteron data.
\end{abstract}

\section{Introduction}
Measurements of the spin-dependent structure function of the 
nucleon, $g_{1N}$, continue to generate enormous interest.  Not only has the 
precision with which the Ellis-Jaffe sum-rule is known improved, but with 
the addition of neutron data from SLAC and SMC the Bjorken sum-rule  
has also been tested \cite{emc,smc,slac}.  Our aim is to provide a 
necessarily brief review of the situation with respect to the sum-rules.  
(For those interested in more detail we suggest some recent reviews 
\cite {ioffe,bass,jelli,ven}.)  We then argue that the emphasis in future 
should be much more on the shape and momentum dependence of $g_{1N}$.  
There has recently been substantial progress in understanding the
flavour and spin dependence of the parton distributions in terms of 
the quark models familiar from low energy spectroscopy and we 
briefly recall the main ideas.

As Nature has not seen fit to 
give us a free neutron target, nuclear and particle physicists 
must work together in order to extract $g_{1n}$. 
Some very recent progress in the 
treatment of deep-inelastic scattering from the deuteron is reviewed.  
Finally we highlight some important open questions which need 
experimental or theoretical attention.

\section{Tests of Sum Rules}
The spin structure function $g_{1N}$ is determined by the difference in 
the cross section for a polarised, virtual photon to be 
absorbed by a nucleon with its spin anti-parallel and parallel to that 
of the photon
\bge
g_1\propto \sigma_\frac{1}{2}-\sigma_\frac{3}{2}.
\ene
As the photon is absorbed by a spin-$\frac{1}{2}$ quark, in the former 
case the absorbing parton must have spin parallel to that of the nucleon 
$(q^\uparrow )$ while in the latter case it must be anti-parallel 
$(q^\downarrow )$.  It is usual 
to define the net spin (helicity) carried by quarks (and anti-quarks) 
of a given flavour as 
\bge
\Delta u(x,Q^2)=u^\uparrow-u^\downarrow+\overline{u}^\uparrow-
\overline{u}^\downarrow,
\ene
etc.  (Quarks and anti-quarks have the same charge squared and contribute 
with the same sign.)  Thus one easily finds
\bge
g_{1p}(x,Q^2)=\frac{1}{2}\left\{\frac{4}{9}\Delta u(x,Q^2)+
	\frac{1}{9}\Delta d(x,Q^2)+\frac{1}{9}\Delta s(x,Q^2)+\ldots\right\},
\ene
and if we define the isovector, octet and singlet flavour combinations in 
the usual way
\bg
\Delta q_3(x,Q^2)&=&\Delta u(x,Q^2)-\Delta d(x,Q^2), \nn\\
\Delta q_8(x,Q^2)&=&\Delta u(x,Q^2)+\Delta d(x,Q^2)-2\Delta s(x,Q^2), \nn\\
\Delta q_0(x,Q^2)&=&\Delta u(x,Q^2)+\Delta d(x,Q^2)+\Delta s(x,Q^2),
\en
we find
\bge
g_{1p(n)}(x,Q^2)=\pm\frac{1}{12}\Delta q_3+\frac{1}{36}\Delta q_8+
	\frac{1}{9}\Delta q_0.
\ene

Using current algebra the integrals of $\Delta q_3$ and $\Delta q_8$ can 
be related to the axial charges in neutron and hyperon $\beta$-decay
\bg
\Delta q_3&=&\int_0^1dx \Delta q_3(x,Q^2)=1.257\pm 0.003, \nn \\
\Delta q_8&=&\int_0^1dx \Delta q_8(x,Q^2)\stackrel{?}{=}0.59\pm0.02,
\en
where the question mark indicates some residual uncertainty over the use
of flavour SU(3) symmetry.
From the definition in equ.(4) it is clear that the integral of 
$\Delta q_0$ is just the fraction of the spin of the nucleon carried by 
its quarks, $\Sigma$ :
\bge
\Sigma=\int_0^1dx \Delta q_0(x,Q^2).
\ene
Within QCD there are also perturbative corrections which vanish 
logarithmically as $Q^2 \rightarrow \infty$, 
so that the integrals of the proton 
and neutron spin structure functions 
\bge
S^{p(n)}(Q^2)=\int_0^1 dx g_{1p(n)}(x,Q^2),
\ene
become \cite{kod,ar,lv}:
\bg
S^{p(n)}(Q^2)&=&\frac{1}{12}\left(1-\frac{\alpha_s}{\pi}-
	3.58\left(\frac{\alpha_s}{\pi}\right)^2-
	20.2\left(\frac{\alpha_s}{\pi}\right)^3\right)
	\left(\pm\Delta q_3+\frac{\Delta q_8}{3}\right)\nn \\
	&+&\frac{1}{9}\left(1-\frac{\alpha_s}{3\pi}-
	1.1\left(\frac{\alpha_s}{\pi}\right)^2\right)\Sigma,
\en
(for three flavours).

The Bjorken sum-rule involves the difference of the integrals for 
the proton and neutron
\bg
S^{Bj}&=&S^p-S^n=\frac{1}{6}\left(1-\frac{\alpha_s}{\pi}-
	3.6\left(\frac{\alpha_s}{\pi}\right)^2\right)\Delta q_3, \nn \\
	&=&0.185\,\,\,{\rm at}\,\,Q^2=10 {\rm GeV^2}.
\en
This particular sum-rule is extremely important as a failure would 
represent a failure of current algebra for the quarks \cite{iz}. The 
Ellis-Jaffe sum-rule \cite{ej} requires a dynamical assumption, namely 
that the polarisation of the strange sea of the nucleon vanishes, 
$\Delta s = 0$. With this reasonable assumption we find $\Sigma\simeq
\Delta q_8\simeq 60\%$ and hence
\bg
S_{EJ}^p&=&0.171\pm0.004,\,\,(0.161\pm0.004),\nn \\
S_{EJ}^n&&=-0.014\pm0.004,\,\,(-0.010\pm0.004)
\en
at $10 GeV^2$ ($3 GeV^2$).
The latest experimental information was presented in detail at this 
meeting \cite{smc,slac} and therefore we simply state the results.

The original EMC measurement \cite{emc} gave $S^p = 0.126\pm 0.010\pm 0.015$ 
at $10.7 GeV^2$.  This corresponds to $\Sigma$ of only $17\%\pm 9\%\pm 14\%$ 
(12\% to 0($\frac{\alpha_s}{\pi}$)). It is clearly compatible with 
{\it none of the nucleon spin being carried by its quarks}, 
which was the origin of the 
term `spin crisis'.  This led to great interest in alternatives to the 
quark model, such as the Skyrme model \cite{bek}.  Indeed the question 
has been asked whether the quark model would have been taken seriously 
if this had been known in the 60's -- rather than the famous prediction   
$\mu_n/\mu_p= -2/3$. On the other hand, a cautious observer might add that 
the result is only 2-3 standard deviations from the Ellis-Jaffe prediction.

At this meeting new values were presented from the SMC and SLAC groups.  
For the proton, SMC finds $S^p = 0.136\pm 0.011\pm 0.011$ at 10 $GeV^2$, 
while a combined analysis with earlier proton data yields $S^p = 0.145
\pm 0.008\pm 0.011$ \cite{smc}.  The preliminary result reported from 
SLAC E143 was $S^p = 0.129\pm 0.004\pm 0.010$ at $3 GeV^2$, while for 
the deuteron and neutron they found $0.043\pm 0.004\pm 0.004$ and $-0.035
\pm 0.0096\pm 0.011$, respectively \cite{slac}. These values are summarised 
in Table 1.

In all cases the experimental values are significantly below the theoretical 
expectation and it is common to conclude that the two experiments agree 
and that $\Sigma$ is of order one third.  On the other hand, we know that the 
integral of $g_1$ changes sign at $Q^2=0$ (to give the DHG sum-rule 
\cite{dhg}), so that there is a dramatic variation in the sum-rule, due to 
non-perturbative effects, between $Q^2=0$ and 3 $GeV^2$.  Vector meson 
dominance has been used by Ioffe and collaborators to provide a physically 
reasonable, smooth interpolation between the low-$Q^2$ and asymptotic 
regimes \cite{ioffe,iof}.  At 3 and 10 GeV$^2$ this approach suggests a 
correction to the measured sum-rule for the proton (neutron) of 29\% (20\%) 
and 8\% (6\%) respectively.  While there is, as yet, no consensus on 
these values \cite{ho} we find the sign and magnitude very reasonable.  
Applying them to the results quoted above, we find the values for 
the integrals and corresponding spin fractions shown in Table 1.
{\footnotesize
\begin{table}
\begin{tabular}{|c|c|c|c|c|} \hline
Experiment & SMC(p) & SMC(p; all data) & SLAC(p; E143) & 
	SLAC(n; E142) \\ \hline
$\langle Q^2\rangle$ ($GeV^2$) & 10 & 10 & 3 & 3 \\
Measured & $0.136\pm0.011$ & $0.145\pm 0.008$ & 
	$0.129\pm 0.004$ & $-0.035\pm0.0096$ \\
	&$ \pm0.011$ & $\pm0.011$ & $\pm0.010$ & $\pm0.011$ \\
Corrected & $0.147\pm 0.016$ & $0.157\pm0.014$ & $0.166\pm 0.014$ & 
	$-0.042\pm0.015$ \\
$\Sigma$ & $34\pm 14\%$ & $43\pm 13\%$ & $62\pm 13\%$ & $28\pm 14\%$ \\ \hline
\end{tabular}
\caption{Integrals of $g_{1p}$ and $g_{1n}$ with and without a
correction for the
non-perturbative dependence on $Q^2$ following 
refs.\protect\cite{ioffe,iof}. }
\end{table}
}
After correction for non-perturbative effects the experiments are 
consistent with an average value of the spin carried by the quarks 
of $42\pm 14\%$ -- somewhat higher than the value of one third found 
without the correction. We believe that $42\pm 14\%$ is 
probably the best estimate at the present time.  It is roughly one 
standard deviation below the value of 60\% implicit in the Ellis-Jaffe 
sum-rule.  Notice that this expectation is not 100\% for the same reason 
that $g_A$ is not 5/3 -- the lower component of the quark wave function 
is p-wave and the quark spin most often points opposite to the total 
angular momentum.  This is a simple, well understood, relativistic 
correction.  Clearly the data is no longer consistent with $\Sigma=0$ 
and a 1 $\sigma$ deviation is not a crisis.  Nevertheless the physics 
is far from clear and in many ways the serious study of the problem is
just beginning!

Much of the theoretical interest in the Ellis-Jaffe sum-rule, and most 
of the papers, have been concerned with the role of 
the U(1) axial anomaly in the flavour singlet piece, $\Delta q_0(x,Q^2)$.  
Following the work of Efremov and Teryaev and Altarelli and Ross \cite{etar} 
(ETAR) -- see also ref.\cite{ccm} -- it is common to correct the 
naive parton model (NPM) value for $\Delta u, \Delta  d$ and $\Delta s$ by
\bg
\Delta u(x,Q^2) &\rightarrow& \Delta u(x,Q^2)|_{NPM}-\delta(x,Q^2),\nn\\
\Delta d(x,Q^2) &\rightarrow& \Delta d(x,Q^2)|_{NPM}-\delta(x,Q^2),\nn\\
\Delta s(x,Q^2) &\rightarrow& \Delta s(x,Q^2)|_{NPM}-\delta(x,Q^2),
\en
where
\bge
\delta(x,Q^2)=\frac{\alpha_s}{2\pi}
	\int_x^1{\cal C}^{\Delta g}\left(\frac{x}{y},Q^2\right)
	\Delta g(y,Q^2) dy,
\ene
and
\bge
\delta\equiv\int_0^1\delta(x,Q^2)=\frac{\alpha_s}{2\pi}\Delta g.
\ene
Here $\Delta g$ is the fraction of the proton spin carried by gluons.  
In this picture
\bge
\Sigma\rightarrow\Sigma_{NPM}-3\delta,
\label{sigeqn}
\ene
and if (following Ellis and Jaffe) one sets $\Delta s$ to zero in the NPM 
we find $\Delta s \neq 0$ through the anomaly.  
Clearly the residual discrepancy in the Ellis-Jaffe sum-rule could be 
fixed with $\delta\approx 5\%$, which corresponds to $\Delta g\sim 1.2$ 
at $10 GeV^2$, or  $\Delta g \sim 1/2$ at a quark model scale 
($\aleq 1 GeV^2$).  The latter agrees rather too well 
with the simple estimate of $\Delta g$ associated with one-gluon-exchange by 
Brodsky and Schmidt \cite{bs}.  Certainly this result is far more 
reasonable that the values of $\Delta g\sim 5$ required when  $\Sigma$ 
was compatible with zero!

We would like to emphasise that one wins in two ways through 
radiation of polarised gluons.  Firstly, as we have just seen,
the point-like coupling to the quarks in the gluon reduces $\Sigma$.  
Secondly, however, these gluons carry spin away from the quarks.  This 
correction was shown to amount to 5-10\% in $\Sigma$ \cite{mt,hm} but as it 
involves non-perturbative QCD has generally been overlooked by the same 
groups who use equ.(\ref{sigeqn}).

As usual with the anomaly there 
has been considerable controversy.  The separation between $\Delta q|_{NPM}$ 
and $\delta$ is not gauge invariant at the operator level.  Indeed this 
lies at the heart of the anomaly \cite{cre}, which arises precisely because 
one cannot satisfy \underline{both} chiral symmetry \underline{and} 
gauge invariance.  In 
the operator product expansion (OPE) there is no operator corresponding 
to $\Delta g$.  From this a number of authors, notably Bodwin and Qiu, 
have argued that the first moment of ${\cal C}^{\Delta g}$ should 
vanish \cite{bq,jm}.

There has also been much debate over the infrared sensitivity
of the separation of $\Delta g_{NPM}$ and $\delta$ 
[4,5,24-27]. In the end, for the gauge usually used 
in the parton model ($A_+=0$), the correction is probably reliable.  
However this raises another problem, the shape of $g_1(x,Q^2)$.  
As emphasised in ref.\cite{bnt,bt},
the gluonic correction effects $g_1(x,Q^2)$ only at very small x (well 
below 0.1).  This is well below the region where the calculations of 
$g_{1p}(x)$ deviate from the data.  The study of the shape of $g_1(x,Q^2)$, 
and particularly the mechanism for transfering strength from intermediate 
x to low x, is therefore a clear priority for future work.  With 
this in mind, in the next section we present some simple ideas which 
help us to understand the shape of $g_{1p}$ and $g_{1n}$.  
Before closing we should mention that not everyone is convinced that 
the effects of the anomaly are confined to small x, and any hint 
of OZI violation in the purely valence region would be most interesting 
indeed \cite {steve}.

\section{Understanding the Shape of $g_{1p}$ and $g_{1n}$.}
Formally the twist-two parton distributions can be written \cite{twist}
\bge
q^{(2)}(x,\mu^2)=\frac{m}{(2\pi)^3}\sum_n\int d\vec p
	|\langle n\vec p|\psi_+(0)|N\rangle|^2\delta(m(1-x)-p_n^+),
\ene
with
\bge
p_n^+=\sqrt{m_n^2+\vec p^2}+p_z\,\,\,>0.
\ene
Here $|n\vec p\rangle$ are a complete set 
of states of momentum $\vec p$ and rest mass $m_n$ and equ.(16) guarantees 
that the distribution vanishes for $x\geq 1$.  This formal expression has 
been used extensively to calculate parton distribution for the MIT bag 
model by the Adelaide group \cite{st,stl,sst}.  An important conclusion 
of that work is that the dominant contribution (at the quark model 
scale $\mu_-^2$ -- c.f. refs.\cite{pp,thom}) arises from a two-quark 
intermediate state, n=2. For a simple nucleon wave function, with 
three quarks in the 1s-state, the matrix element 
$\langle n=2,\vec p|\psi_+(0)|N\rangle$ should peak near $|\vec p|=0$.  
Hence the maximum value of $q^{(2)}(x,\mu^2)$ should occur at 
$x\sim\frac{m-m_2}{m}\sim(\frac{1}{3}-\frac{1}{4})$ -- from the   
$\delta$-function -- and the dominant contribution at large x
will be the one associated with the lowest value of $m_2$.  This was 
first realised by Close and Thomas who stressed the importance for both 
the flavour and the spin dependence of the parton distribution \cite{ct}.  
For example, using the standard SU(6) wave function for the proton 
one easily finds that a valence d-quark has only an S=1 pair 
as spectator, while a u-quark has a 50-50 chance of S=0 or S=1.
Whatever mechanism is used to generate the N-$\Delta$ mass splitting 
-- e.g. gluon exchange, pion exchange or instantons -- gives typically 
$m_2(S=1)-m_2(S=0)\approx200MeV$. Hence the u-quarks should dominate at 
large x -- i.e. one expects $d/u \ll 1$ as $x\rightarrow 1$, as found 
experimentally.

For a polarised proton (neutron) only the valence $u^\uparrow 
(d^\uparrow)$ is found with an $S=0$ pair and hence $u^\uparrow 
(d^\uparrow)$ should dominate at large x.  From equ.(2) we see that 
$g_{1p}$ must be positive, as must $g_{1n}$, at intermediate and large x.  
Note that all this follows from the {\it same physics} 
required in low energy spectroscopy (and a simple SU(6) 
wave function).  It works because deep inelastic scattering involves a 
light-cone correlation function and the energy of the quark is as 
important as its momentum.  Of course, a full calculation should also 
include the configuration mixing in the wave function, but this is not
the {\it leading} effect. In passing we 
note that this same approach to calculating parton distributions leads
to an unexpectedly large violation of charge symmetry for the minority
valence quarks ($d^{(p)}\neq~u^{(n)}$) \cite{sat,rtl}.
%
\begin{figure}[htb]
\centering{\ \psfig{file=nwrevn.ps,height=9cm}}
\caption{ Data for $g_{1n}$ from SMC and SLAC in comparison with
predictions based on the bag model with (dashed) and without (solid) a
phenomenological correction attributed to the axial anomaly -- from
ref.\protect\cite{anom}.}
\label{g1n}
\end{figure}
%

It is somewhat disappointing that the focus on sum-rules means that we 
still do not know whether $g_{1n}$ does become positive at intermediate and 
large x.  Although the approach just described is unambiguous, the 
inclusion of the correction associated with the axial anomaly could change 
this and it is very important to know.  Figure 1 illustrates this by 
showing the prediction for $g_{1n}$ from the bag model (with a bag radius 
R=0.6 fm) \cite{sst}, with and without a phenomenological singlet 
term added to fit the proton data \cite {anom}.  The dashed curve was 
then a prediction for $g_{1n}$ published before the data. Recall 
that the earlier phenomenology of Carlitz-Kaur \cite {ck} and Sch\"afer 
\cite {sch} used counting rules to parameterize the effect of 
admixing polarised gluons in the proton wave function.  As we have already 
noted, this complements the work of ETAR - see also refs.\cite{mt,hm}.  
It would be desirable to have a full microscopic calculation, 
including this effect and the Close-Thomas mechanism, within a model 
which is consistent with low energy hadronic properties.

As a final consideration in calculating the shape of the spin
distributions we must also mention the role of chiral symmetry. Chiral
quark models, like the cloudy bag\cite{cbm}, have proven very successful
in dealing with low-energy hadronic properties. The charge form-factor
of the neutron, $G_{En}$, is a famous example, where the positive core
and negative tail of the corresponding charge density is a first-order
effect in the perturbative dressing of the nucleon bag -- through the
process $n \rightarrow p_{Bag} \pi^-$\cite{cbm,neut}. (Indeed, within
the cloudy bag model an accurate measurement of the zero in
$\rho^n_{ch}(r)$ would determine the radius within which the quarks are
confined.) For the spin problem the role of the pion cloud is more
complicated. For the proton the dominant, pionic correction is
$p^\uparrow \rightarrow n^\downarrow \pi^+(l_z=+1)$, while for the
neutron it is $n^\uparrow \rightarrow p^\downarrow \pi^-(l_z=+1)$.
Combined with the reduction in the bare nucleon contribution through the
wave function renormalisation constant, Z, both of these effects tend to
reduce the positive values of $g_1$ at large x \cite{schreib}. The only
complete investigation of pionic corrections has been carried out for
the Ellis-Jaffe sum-rule \cite{pion}, where it gives a small reduction 
($O(5\%)$). This reduction would be larger were it not for an $N-\Delta$
interference term. The x-dependence of the latter is a completely new
term which has never been calculated.

\section{Nuclear Spin Structure Functions:}
In order to study quark-parton distributions in the neutron, one 
must deal with either 
the deuteron or some heavier nucleus.  In the spin-dependent case the 
two nuclei studied are polarised deuterium ($\vec D$) and {$^3He$}.  
The simplest theoretical approach for deuterium is to approximate it 
as a free neutron and proton with polarisation ($1-\frac{3}{2} \omega_D$) 
- where $\omega_D$ is the deuteron d-state probability.  At the next level 
of sophistication we have a convolution of the free nucleon structure 
function with the light-cone 
momentum distribution of nucleons in the nucleus \cite {con,cio}:
\bge
g_{1D}(x,Q^2)=\sum_{N=n,p}\int_x \frac{dy}{y}\Delta f_{N/D}(y) 
	g_{1N}(\frac{x}{y},Q^2).
\ene

Strangely the situation for polarised structure functions is better than 
for the unpolarised case, where data is still often analysed \cite{nod} 
using ``smearing functions" pre-dating \cite{fs} the original 
EMC effect \cite{emco} - and therefore ignoring binding 
corrections \cite{bick,mts1,mts2} (see however ref.\cite{umnik}). 
Tokarev found that for the 
polarised case the convolution gave a result very close to that 
obtained with a constant depolarisation for $x\aleq0.7$ \cite{tok} 
- see also Kaptari and Umnikov \cite{ku}.

Reassuring as these results may 
be, the only way to test the reliability of the convolution is to 
do better.  Even within the impulse approximation (IA) one may ask, 
for example, whether off-shell effects might spoil the convolution form.  
For the unpolarised case this was studied by Melnitchouk, Thomas and 
Schreiber \cite{mts1,mts2} who found that even in the Bjorken limit 
the Dirac structure of the nucleon involves not one but three functions
\bge
\hat{W}^{\mu\nu}=P_T^{\mu\nu}(\chi_0(p,q)+\pslash\chi_1(p,q)+\qsl\chi_2(p,q)).
\ene
Thus, whereas the free structure function involves one linear combination
\bge
F_2^{free}\propto m\chi_0+m^2\chi_1+p\cdot q\chi_2,
\ene
obtained by tracing $\hat{W}^{\mu\nu}$ with $(\pslash+m)$, in a nucleus, 
where the nucleon propagator is $({\not\!\!A}+B)$, we find
\bge
F_2^{bound}\propto B\chi_0+A\cdot p\chi_1+A\cdot q\chi_2,
\ene
which is \underline{not} proportional to $F_2^{Free}$.

To study the practical significance of this breakdown, one needs a 
covariant description of the $\vec D\rightarrow \vec N N$ vertex 
as well as the off-shell nucleon structure function.  
For the former we can use the vertex of refs.\cite{gross}.  
For the latter Melnitchouk, Piller and Thomas \cite{mpt} have 
employed a covariant version of the physics presented in sect. 3.  
That is, they constructed a set of phenomenological functions for the 
$N\rightarrow q$ (di-quark) vertices which were fitted to free nucleon 
data.  As before, the di-quark states were either scalar or pseudovector.  
After considerable Dirac algebra one finds that the anti-symmetric 
nucleon tensor, surviving for massless quarks, in the Bjorken limit is
\bge
\hat{W}_{A-S}^{\mu\nu}=i {\epsilon}^{\mu\nu\alpha\beta}
	\left\{q_\alpha p_\beta \pslash\gamma_5 G_{(p)}
	+q_\alpha p_\beta \qsl G_{(q)}
	+q_\alpha\gamma_\beta\gamma_5 G_{(\gamma)}\right\},
\ene
so that
\bge
g_{1N}^{Free}\propto p\cdot q\left[p\cdot q G_{(q)}+G_{(\gamma)}\right].
\ene
For the deuteron case one must trace $\hat{W}_{A-S}^{\mu\nu}$  with a 
term involving 
$\gamma_5\gamma_\lambda{\cal A}(P, S, p)$,
with the result:
\bge
g_{1D}(x)\propto\sum_{N=n,p}\int dy dp^2\left\{{\cal A}\cdot q
	\left[p\cdot q G_{(q)}^N+G_{(\gamma)}^N\right]
	+p\cdot q {\cal A}\cdot p G_{(p)}^N\right\}.
\ene
The last term is not present in the free case, so that in principle 
the convolution must fail.
\begin{figure}[htb]
\centering{\ \psfig{file=spinlet2.ps,height=9cm}}
\caption{Ratio of deuteron and isoscalar nucleon structure functions in the
full model (solid) compared with a constant depolarization factor
$1 - 3/2 \omega_D$ (dotted, for $\omega_D = 4.7\% $\protect\cite{gross}) .
Also shown (dashed line) is the ratio of $g_{1D}$ calculated via convolution
and in the full model -- from ref.\protect\cite{mpt}.}
\label{mel}
\end{figure}

Figure 2 shows the numerical results obtained by Melnitchouk et al. 
\cite{mpt} for the ratio of the deuteron to isoscalar nucleon spin structure 
function.  The full calculation (solid line) is very close to the 
simple approximation of a constant depolarisation (dotted line) for 
$x\aleq 0.7$.  This agrees with the result found in earlier work using 
the convolution approach.  Indeed we see from the dashed curve that an 
appropriate convolution  is very close to the full off-shell calculation 
for all $x\aleq 0.9$.  In practice the additional term in equ.(24) 
is of order $(v/c)^3$ and negligible for the deuteron.  In the 
end the major uncertainty in extracting the neutron structure function 
is not the failure of convolution or even the assumption of a constant 
depolarisation, but the old chestnut of the lack of knowledge of 
the deuteron d-state probability.  This ranges between 4 and 7\%, with 
modern, non-local potentials preferring $5\pm1\%$.  At $x = 0.1$, for 
example, a variation of 1\% in $\omega_D$ corresponds to an error of roughly 
10\% in $g_{1n}$ \cite{mpt}.  At present this is smaller than the 
experimental errors but will soon be significant.

To conclude this discussion, 
we must note that the IA is by no means secure.  For heavier nuclei 
it has been shown to significantly overestimate the effect of binding 
because deep-inelastic scattering determines the energy and momentum 
distribution of quarks \underline{not} nucleons \cite{thomas,ks}!  
This has not yet been investigated for the deuteron where the IA may not 
be so bad.  Theoretical studies of $^3\vec{He}$ have so far not gone 
beyond an essentially non-relativistic convolution because of the absence 
of a relativistic model of the necessary vertices.  In view of the deuteron
results it would be surprising if there were a major problem, nevertheless 
it should be checked.

Finally we note that the actual structure of the bound nucleon may change.  
For the unpolarised case this has been shown to be a very small effect 
\cite{thomas} within the Guichon model \cite{guich}.  However, spin 
structure functions involve the lower components of the quark wave 
functions more directly and these change most in nuclear matter.  
It has been suggested in ref.\cite{ku}, for example, that 
$(g_A/g_V)_{Bound}$ differs from the free value by about 5\%.  
It is therefore an urgent matter to check this issue.

\section{Concluding Remarks}
As we have seen there has been remarkable progress in the 
experimental determination of the spin structure functions of the 
proton and the neutron.  The Bjorken sum-rule has been verified at 
the 10\% level (at 1$\sigma$).  In view of its significance 
it is very important to continue to reduce this error.  With 
respect to the Ellis-Jaffe sum-rule it is now clear that there is 
no longer a `crisis' in the sense that none of the nucleon spin is 
carried by quarks.  With a reasonable estimate of the non-perturbative
corrections we find that all experiments are consistent with 
$\Sigma\sim 42\pm 14\%$ compared with the expectation of about 60\% 
of the nucleon spin carried by quarks.

However, {\it the end of the crisis is the beginning of some very 
challenging and fascinating physics}.  It is now known that the 
Regge behaviour usually assumed in extrapolating to $x=0$ is not 
appropriate to the flavour singlet component, which may have 
logarithmic corrections \cite{cr}.  Bass and Landshoff suggest these 
might contribute as much as 25\% to $\Sigma$  \cite{bl}, while Close and
Roberts have suggested recently that the error might be even larger\cite{ral}.
There is as yet no consensus on the size of 
the higher twist corrections. Both of these issues need thorough 
experimental study of the $x$ and $Q^2$ variation of the structure 
functions as well as more theoretical work.

The shape of the $x$ distribution of the neutron is still not well 
known.  In view of the beautiful link between low energy hadron 
structure and parton distributions outlined in Sect. 3 it is very 
important to determine whether $g_{1n}$ does become positive at intermediate 
and large $x$. Related to this is the question of whether strength 
is shifted from intermediate $x$ to the region below 0.1, and if so what 
mechanism is responsible.  The role of the axial anomaly is particularly 
fundamental, and can in principle be determined by comparing $g_1$ with 
a C-odd structure function like $g_3$ \cite{bass}.

The use of light nuclear targets to extract information on the neutron 
seems to be well under control.  We have seen that an appropriate 
convolution formula is very accurate when compared with a full off-shell 
calculation.  The major problem for the deuteron is the uncertainty 
over its d-state probability.  This may also be a problem for $^3He$.  
Shadowing and meson exchange corrections have been investigated in detail 
for the unpolarised case \cite{shad} but much more must be done for 
the polarised case \cite{kap,hood}.  There is so far no estimate of the 
effect on $g_1$ of a possible change in the internal nucleon structure.

Throughout our discussion we have followed the conventional 
approach of ignoring charm quarks (i.e. $\Delta c$ is set to zero). Since 
most data points are actually above charm threshold, this needs more 
investigation \cite {ell,basst}.  A related problem concerns the 
suggestion of using neutrino proton elastic scattering to measure 
$\Delta s$.  As the neutral current couples to a non-singlet axial 
current it is less than obvious that such an experiment can yield 
information on a singlet quality like $\delta$. Thus one may learn 
about $\Delta s|_{NPM}$ rather than $\Delta s$ \cite{anom,basst}.

The breadth of open questions in this field is enormous - ranging from 
issues of quark structure in nuclei, to Pomerons and the axial anomaly.  
It is truly an appropriate meeting ground for nuclear and particle 
physicists and an appropriate beginning to the first joint Spin 
Conference.  I am grateful to the Organising Committee(s) for 
the invitation to present these thoughts.

\section*{Acknowledgments}
It is a pleasure to thank my colleagues, particularly S. Bass, 
W. Melnitchouk, G. Piller, A. Schreiber and F. Steffens,  
who have taught me a great deal about the matters presented here.  
I would also like to thank A. Rawlinson and A. Shaw for their assistance
in the production of this manuscript.
This work was supported by the Australian Research Council.


%
\end{document}